\documentclass[referee,sn-aps]{sn-jnl}

\usepackage[normalem]{ulem}
\usepackage{graphicx}%
\usepackage{multirow}%
\usepackage{amsmath,amssymb,amsfonts}%
\usepackage{amsthm}%
\usepackage{mathrsfs}%
\usepackage[title]{appendix}%
\usepackage{xcolor}%
\usepackage{textcomp}%
\usepackage{manyfoot}%
\usepackage{booktabs}%
\usepackage{algorithm}%
\usepackage{algorithmicx}%
\usepackage{algpseudocode}%
\usepackage{listings}%
\usepackage{amsthm}
\usepackage[margin=3.0cm]{geometry}

\newcommand{\be}{\begin{equation}}
	\newcommand{\ee}{\end{equation}}
\newcommand{\bea}{\begin{eqnarray}}
	\newcommand{\eea}{\end{eqnarray}}

\newcommand{\comment}[1]{}

\theoremstyle{thmstyleone}%
\theoremstyle{thmstyletwo}%

\theoremstyle{thmstylethree}%

\begin{document}

\title[Domain growth]{Domain Growth in Long-range Ising Models with Disorder}


\author[1]{\fnm{Ramgopal} \sur{Agrawal}}\email{ramgopal.sps@gmail.com}
\author[2]{\fnm{Federico} \sur{Corberi}}\email{corberi@sa.infn.it}
\author[3]{\fnm{Eugenio} \sur{Lippiello}}\email{eugenio.lippiello@unicampania.it}
\author*[4]{\fnm{Sanjay} \sur{Puri}}\email{purijnu@gmail.com}

\affil[1]{\orgdiv{Dipartimento di Fisica}, \orgname{Sapienza Universit\`a di Roma}, \orgaddress{\street{Piazzale Aldo Moro 5}, \city{Rome}, \postcode{00185}, \country{Italy}}}

\affil[2]{\orgdiv{Dipartimento di Fisica ``E. R. Caianiello'', and INFN, Gruppo Collegato di Salerno, and CNISM}, \orgname{Universit\`a  di Salerno}, \orgaddress{\street{via Giovanni Paolo II 132}, \city{Fisciano (SA)}, \postcode{84084}, \country{Italy}}}

\affil[3]{\orgdiv{Dipartimento di Matematica e Fisica}, \orgname{Universit\`a della Campania}, \orgaddress{\street{Viale Lincoln 5}, \city{Caserta}, \postcode{81100}, \country{Italy}}}

\affil*[4]{\orgdiv{School of Physical Sciences}, \orgname{Jawaharlal Nehru University}, \orgaddress{\street{New Mehrauli Road}, \city{New Delhi}, \postcode{110067}, \country{India}}}


\abstract{Recent advances have highlighted the rich low-temperature kinetics of the long-range Ising model (LRIM). This study investigates domain growth in an LRIM with quenched disorder, following a deep low-temperature quench. Specifically, we consider an Ising model with interactions that decay as $J(r) \sim r^{-(D+\sigma)}$, where $D$ is the spatial dimension and $\sigma > 0$ is the power-law exponent. The quenched disorder is introduced via random pinning fields at each lattice site. For nearest-neighbor models, we expect that domain growth during activated dynamics is \textit{logarithmic} in nature: $R(t) \sim (\ln t)^{\alpha}$, with growth exponent $\alpha >0$. Here, we examine how long-range interactions influence domain growth with disorder in dimensions $D = 1$ and $D = 2$. In $D = 1$, logarithmic growth is found to persist for various $\sigma > 0$. However, in $D = 2$, the dynamics is more complex due to the non-trivial interplay between extended interactions, disorder, and thermal fluctuations.}




\maketitle

\section{Introduction}
\label{s1}

\begin{figure}[t!]
	\centering
	\includegraphics[width=0.95\linewidth]{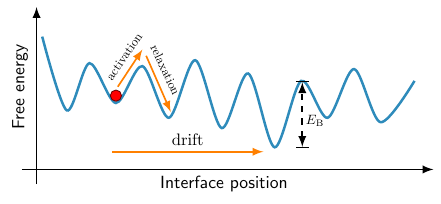}
	\caption{Schematic of interface-activated-dynamics (red circle) in a quenched disorder landscape with local energy minima of height $E_{\rm B}$. Thermal activation enables the domain wall to overcome energy barriers, followed by relaxation into new minima. A ferromagnetic long-range interaction induces a drift, assisting the interface in escaping minima.}
	\label{fig1}
\end{figure}

Phase ordering, or domain growth, has been a topic of intense research over the past 40 years~\cite{Bray94,dp04,PuriWad09,CRPHYS_2015__16_3_332_0}. The central feature of domain growth is the rampant growth of domains followed by dynamical scaling in terms of a \textit{characteristic} length scale $R(t)$. A paradigmatic example is the nonequilibrium evolution of an Ising ferromagnet after a quench from the paramagnetic to the ferromagnetic phase. During this process, domains of aligned spins form and grow via interface diffusion, obeying (for sufficiently short-range interactions) the conventional Lifshitz-Allen-Cahn (LAC) law, $R(t) \sim t^{1/2}$ \cite{AllenCahn79,PhysRevB.37.9481}. The introduction of quenched disorder~\cite{PhysRevLett.54.2708,Puri2004,Paul_2004} pins domain interfaces at energetically favorable positions, leading to activated dynamics (see Fig.~\ref{fig1}). In this regime, interfacial dynamics is no longer diffusive but instead depends on disorder-induced energy barriers. Within the Huse-Henley framework~\cite{PhysRevLett.54.2708} and subsequent studies (discussed later), it is now well established that the domain growth law during the activated processes is logarithmic in nature, $R(t) \sim (\ln t)^{\alpha}$; with $\alpha > 0$ is the growth exponent. These problems are well understood in systems with short-range interactions, such as Ising models with nearest-neighbor (NN) pairwise couplings, where disorder can be introduced in various ways.

In particular, the random-field Ising model (RFIM) represents a canonical example of short-range disordered systems, where quenched disorder is introduced via site-dependent local random fields. As one of the most paradigmatic models of disorder, the RFIM has been extensively studied both for its theoretical richness \cite{PhysRevLett.35.1399,Young_1977,PhysRevLett.43.744,doi:10.1142/9789812819437_0009,BELANGER1991272,fytas2018review} and its diverse applications, ranging from glass-forming liquids~\cite{PhysRevLett.112.175701,PhysRevE.102.042129} to computer vision~\cite{56204,969114}. In these systems, the on-site disorder significantly alters phase transitions and causes complex dynamics, including slow domain growth, aging, and glass-like behavior \cite{doi:10.1142/3517,CorCugYos11}. For the one-dimensional (1D) RFIM, interfacial dynamics can be mapped to a Sinai-type random walker in a force field~\cite{doi:10.1137/1127028,BOUCHAUD1990285}, leading to logarithmic growth~\cite{PhysRevE.64.066107,PhysRevE.65.046114}. In addition, in dimension $D \geq 2$, the asymptotic growth law has long been found to be logarithmic \cite{Puri_1993,PhysRevLett.71.3501,Oguz_1994,Aron_2008,corberi2012crossover}.

The situation becomes even more complex when the slowdown induced by the quenched disorder competes with the acceleration typically caused by long-range interactions. This interplay is the central focus of the present review, which examines random field models with long-range interactions, commonly referred to as the \textit{random-field long-range Ising model} (RFLRIM; see Sec.~\ref{s2} for model details). These systems are poorly understood, and only recent advances in pure long-range Ising models (LRIM)~\cite{corberi2019one,Corberi_2019,CMJ19,PhysRevLett.125.180601,PhysRevE.103.012108,PhysRevE.103.052122,PhysRevE.105.034131,Corberi_2024} have revealed their richness. In the following, we will consider the case where the exchange coupling between spins decays spatially as a power law, $J(r) \sim r^{-(D+\sigma)}$, where $\sigma$ is the power-law exponent. Domain growth in these systems exhibits a particularly complex and rich phenomenology. Even in pure LRIMs, multiple growth regimes emerge~\cite{corberi2019one,PhysRevE.103.012108,PhysRevE.103.052122}, governed by the competition between long-range-induced drift and thermal diffusion. These regimes maintain purely algebraic growth laws, $R(t) \sim t^{1/\bar{z}}$, with the exponent $\bar{z}$ depending on both $\sigma$ and temperature (as discussed in detail later). In the presence of quenched disorder, the interfacial dynamics becomes both drifted and activated (Fig.~\ref{fig1}). This raises important questions: Does the Huse-Henley argument for logarithmic growth \cite{PhysRevLett.54.2708} remain valid in disordered long-range systems? If so, what role does long-range-induced drift play? These questions remain largely unexplored in the literature, except our recent study~\cite{PhysRevE.108.044131} where domain growth in a 1D RFLRIM is investigated.

Following a pedagogical approach, this paper first discusses domain growth in 1D systems with \textit{Glauber} nonconserved kinetics and then advances the concepts to $2D$ systems. To focus our study, we restrict ourselves to the weak long-range limit ($\sigma > 0$), as the strong limit $(\sigma \le 0)$ introduces additional complexities, such as the loss of additivity and extensivity \cite{campa2014physics,CAMPA200957,CorIanKumLipPol21}. Through this work, our aim is to elucidate the rich phenomenology of domain growth and dynamical scaling in disordered systems with long-range interactions.

The paper is structured as follows. In Section~\ref{s2}, we introduce the model (RFLRIM) and discuss the relevant literature. Section~\ref{s3} details our numerical approach, including simulation methods and measured observables. We present our main results for $1D$ systems in Section~\ref{s4}, followed by the $2D$ case in Section~\ref{s5}. Finally, Section~\ref{s6} provides a comprehensive discussion of our findings and outlines potential future research directions.

\section{The Model}
\label{s2}

The Hamiltonian of the RFLRIM is given by
\be
{\cal H}(\{s_i\})= -\sum_{j<i} J(r_{ij}) s_i s_j - \sum _{i=1}^{N} h_i s_i, \quad s_i = \pm 1
\; ,
\label{ham}
\ee
where $N = L^{D}$ is the number of sites on a $D$-dimensional square lattice, and $L$ is the linear size of the system. The exchange coupling $J(r)$, which defines the ferromagnetic interaction between spins $s_i$ and $s_j$ lying at distance $r_{ij} =\vert i - j \vert$, takes the following form
\be
J(r) = \frac{J_0}{r^{D+\sigma}}
\; ,
\label{int}
\ee
where the power law exponent $\sigma$ is set to $\sigma >0$, and the constant $J_0>0$ denotes the strength which is set to unity without loss of any generality. The NN Ising model is recovered when $\sigma \longrightarrow \infty$, that is, $J(r) = J_0 \delta_{r,1}$. The field variable $h_i$ in Eq.~\eqref{ham} represents the on-site random field, which is drawn from a Gaussian distribution
\be
P(h_i) = \frac{1}{\sqrt{2 \pi \Delta^2}} e^{-h_i^2/(2\Delta^2)}
\; ,
\ee
where the standard deviation of the distribution $\Delta$ controls the strength of the disorder.

The phase transitions in the random field models with LR interactions are scarcely understood. First, let us recall the well-established results for systems without LR interactions, where the lower critical dimension $D_{\ell} = 2$ was initially predicted by Imry and Ma~\cite{PhysRevLett.35.1399} through domain-wall stability arguments and by Young~\cite{Young_1977} using perturbative theory. For dimensions $D > D_{\ell}$, a stable ferromagnetic phase can exist at non-zero temperatures. The existence of phase transitions in $D \ge 3$ for low temperatures and sufficiently weak disorder was rigorously demonstrated by Bricmont and Kupiainen~\cite{PhysRevLett.59.1829} and Aizenman-Wehr~\cite{PhysRevLett.62.2503,aizenman1990rounding}. More recently, Ding and Zhuang~\cite{https://doi.org/10.1002/cpa.22127} provided a simpler proof of phase transitions in $D \ge 3$ using Peierls' argument~\cite{Peierls_1936}. Extending the Ding-Zhuang approach to LR systems, Affonso et al.~\cite{affonso2024} have made progress in understanding phase transitions in higher-dimensional RFLRIM. They showed that for $\sigma > 0$, there is a phase transition at $T_c \ne 0$ in $D \ge 3$ when the disorder is sufficiently weak.

For lower-dimensional RFLRIM ($D = 1,2$), Aizenman and Wehr \cite{PhysRevLett.62.2503,aizenman1990rounding} demonstrated the uniqueness of the Gibbs measure when $\sigma > D/2$ in dimensions $D \leq 2$. However, this left open the crucial question: Does long-range order (LRO) exist for $\sigma < D/2$? In the $D=1$ case, Okuyama and Ohzeki~\cite{okuyama2025existence} studied this model on Dyson hierarchical lattices with decaying interactions as in Eq.~\eqref{int}. Their rigorous proof established that for $0 < \sigma < 1/2$, LRO persists at finite low temperatures (including $T=0$) when the random field strength is sufficiently weak. This result confirms the Aizenman-Wehr threshold value $\sigma_{\rm AW} = 1/2$ in one dimension. Although there is no rigorous proof for $D=2$, physical intuition suggests that LRO can persist similarly when $\sigma < \sigma_{\rm AW}(=1)$, provided that the strength of the disorder remains sufficiently weak. This expectation comes from the analogous behavior observed in the $D=1$ case and the general trend of increased ordering stability in higher dimensions.

In this work, we focus on domain growth in the RFLRIM for dimensions $D=1,2$ following a temperature quench from the high-temperature paramagnetic phase to low-temperature regimes. Our study employs numerical simulations of Glauber kinetics (with non-conserved order parameter) to characterize the nonequilibrium evolution of the system. 

The following section discusses our numerical methodology, including simulation details and key observables used to quantify the domain growth kinetics.

\section{Numerical Details and Observables}
\label{s3}

Since the model system~\eqref{ham} does not have any intrinsic dynamics, we evolve it using Glauber spin-flip kinetics \cite{doi:10.1063/1.1703954}. We use Metropolis transition rates at the temperature $T$:
\be
w(s_i\to -s_i) = N^{-1}\min \left( 1,e^{-\Delta E/(k_{\rm B}T)} \right)
\; ,
\label{metrop}
\ee
where $\Delta E$ is the energy difference in the proposed spin-flip move, and $k_{\rm B}$ is the Boltzmann constant. The simulation time is measured in terms of Monte Carlo steps (MCS), each comprising $N$ elementary spin-flip attempts. The system is initialized at time $t=0$ in a random configuration sampled from a $T=\infty$ equilibrium ensemble, representing a fully disordered paramagnetic phase. We then simulate the time evolution at temperature $T$ by applying the transition rates given in Eq.~\eqref{metrop}.

As we are dealing with spatially extended interactions~\eqref{int}, the finite size effects become more relevant. Therefore, we implement periodic boundary conditions (PBCs) in all spatial directions. However, in systems with long-range interactions, a \textit{trivial} minimum-image convention approach does not work. For this purpose, we exploit a sophisticated approach~\cite{fs2002}, where the \textit{effective} interaction between two spins $s_i$ and $s_j$ on a lattice is expressed as an infinite summation over all replicas lying across the periodic boundaries:
\be
\label{jsum}
J^*(s_i,s_j) = \sum_{\vec n} \frac{1}{\vert \vec n L + \vec r_{i} - \vec r_{j} \vert ^{D+\sigma}}
\; ,
\ee
where the summation is over the displacement vector $\vec n =  (n_1, n_2,\ldots, n_k,\ldots)$ with $n_k = 0, \pm 1, \pm 2, \ldots$ representing the coordinates of various replicas. The original lattice is located at the origin ($n_k = 0$). In the dimensions $D > 1$, the above summation is dealt with numerically by adapting the Ewald summation method~\cite{fs2002,PhysRevE.95.012143}, which exploits a clever trick to split the above slow converging summation~\eqref{jsum} into two rapidly converging summations (this allows to truncate the summations with least error bounds); see Ref.~\cite{PhysRevE.103.012108} for more technical details of the method. Except in $D=1$, the infinite summation in Eq.~\eqref{jsum} can be analytically solved~\cite{PhysRevB.86.014431,PhysRevE.108.044131} without any truncation using Hurwitz zeta functions $H(a,b)$~\cite{abramowitz1988handbook} as
\be
\label{jsum_2}
J^*(s_i,s_j) = \frac{1}{L^{1+\sigma}} \left[ H\left( 1+\sigma, \frac{r_{ij}}{L} \right) + H\left( 1+\sigma, 1 - \frac{r_{ij}}{L} \right) \right]
\; ,
\ee
where $r_{ij}$ denotes the distance between spins $s_i$ and $s_j$ on the original lattice.

During domain growth kinetics, the time-dependent length scale $R(t)$ can be calculated from the inverse of defect density $\rho(t)$~\cite{Bray94}. In $D=1$, this definition of $R(t)$ is appropriate as the defect density $\rho(t)$ is easy to calculate (the number of misaligned spins divided by the total number of spins $N$). However, in higher dimensions, $R(t)$ is usually extracted from the equal-time spatial correlation function $C(r,t)$,
\be
\label{corr}
C(r ,t) = \frac{1}{N} \sum_{i=1}^N \left[\overline{\left<s_{i}(t) s_{i+r}(t)\right>} - \overline{\left<s_{i}(t)\right>} ~ \overline{\left<s_{i+r}(t)\right>}\right]
\; ,
\ee
where the notation $\overline{\left <\ldots \right >}$ denotes the average over various disorder realizations and random initial conditions. $C(r,t)$ shows dynamical scaling with respect to the time-dependent length scale $R(t)$,
\be
\label{cf_scale}
C(r,t) \equiv f\left(\frac{r}{R(t)}\right)
\; .
\ee
Here, $f(x)$ is a scaling function. We define $R(t)$ as the distance over which $C(r,t)$ decays to $0.5$ of its maximum value, that is, $C(L,t)=0.5 \times C(0,t)$.
The term super-universality (SU) is used to indicate that all dependence on the model parameters is contained in the growth law, while the scaling function $f(x)$ is universal. This implies that once the distances are rescaled by $R(t)$, the form of the correlation function or other observables becomes independent of the microscopic details of the model or the quench protocol.

To quantitatively study the growth laws in the presence of quenched disorder, we calculate the \textit{effective} dynamic exponent $z_{\rm eff}$ as follows:
\be
\label{growth_exp}
z_{\rm eff} (t)=\left [\frac{d \ln R (t)}{d \ln t}\right]^{-1}
\; .
\ee
Note that both the characteristic length scale $R(t)$ and effective exponent $z_{\rm eff}(t)$ depend on the disorder strength $\Delta$, though we omit this dependence in our notation for simplicity. The behavior of $z_{\rm eff}(t)$ distinguishes different growth regimes:
\begin{itemize}
	\item For power-law growth [$R(t) \sim t^{1/\bar{z}}$], $z_{\rm eff}(t)$ converges to a constant value $\bar{z}$.
	\item For logarithmic growth [$R(t) \sim (\ln t)^{\alpha}$], $z_{\rm eff}(t)$ shows a smooth upward trend with time.
\end{itemize}

In disordered systems, we typically observe multiple growth regimes separated by crossover regions, reflecting the competition between different dynamical mechanisms.

\section{Domain Growth in One Dimension}
\label{s4}

This section presents our numerical results for the 1D RFLRIM, based on our recent investigation in Ref.~\cite{PhysRevE.108.044131}. To achieve reliable statistical accuracy, we average all measured quantities over 50-100 independent simulation runs, with each run employing distinct initial condition and random field realization. To avoid finite size effects, a large system of size $N (=L) = 2^{20}$ is considered.

Let us start by discussing the domain growth law. For the convenience of the reader, we first briefly recap the growth laws in a pure LRIM. Since the LR interaction introduces a drift on the domain interfaces, the growth law changes a lot with respect to the free-diffusive dynamics of the NN model. Bray and Rutenberg~\cite{PhysRevE.49.R27,PhysRevE.50.1900} had predicted the modified growth law using a continuum description, which were recently re-derived by Corberi et al.~\cite{corberi2019one} for the discrete Ising model. The Bray-Rutenberg (BR) predictions for a quench to non-zero temperature are as follows,
\bea
R(t) &\sim& t^{1/(1+\sigma)}
\; , \quad \sigma < 1
\; , \nonumber \\
&\sim& t^{1/2}
\; , \quad \sigma > 1
\; .
\label{BR}
\eea
The growth law for the pure NN Ising model is $R(t) \sim t^{1/2}$, which is clearly obtained from the BR predictions when setting $\sigma \rightarrow \infty$. In addition to the above law~\eqref{BR}, Corberi et al.~\cite{corberi2019one} found that, for discrete spin variables, there are several pre-asymptotic growth regimes depending on the value of $\sigma$. For $\sigma < 1$, a ballistic regime, $R(t) \sim t$, occurs on early timescales. However, for $\sigma > 1$, there is also a slower regime right between the ballistic and asymptotic BR regimes, where $R(t) \sim t^{1/(1+\sigma)}$. The crossover length for ballistic regime is,
\be
R(t) \simeq R_b \sim \left (\frac{2}{\sigma T}\right) ^{1/\sigma}
\; ,
\ee
while for $\sigma > 1$ the asymptotic growth [$R(t) \sim t^{1/2}$] begins when
\be
R(t) \simeq R_d \sim \left (\frac{2}{\sigma T}\right) ^{1/(\sigma-1)}
\; .
\ee

\begin{figure}[t]
	\centering
	\includegraphics[width=0.95\linewidth]{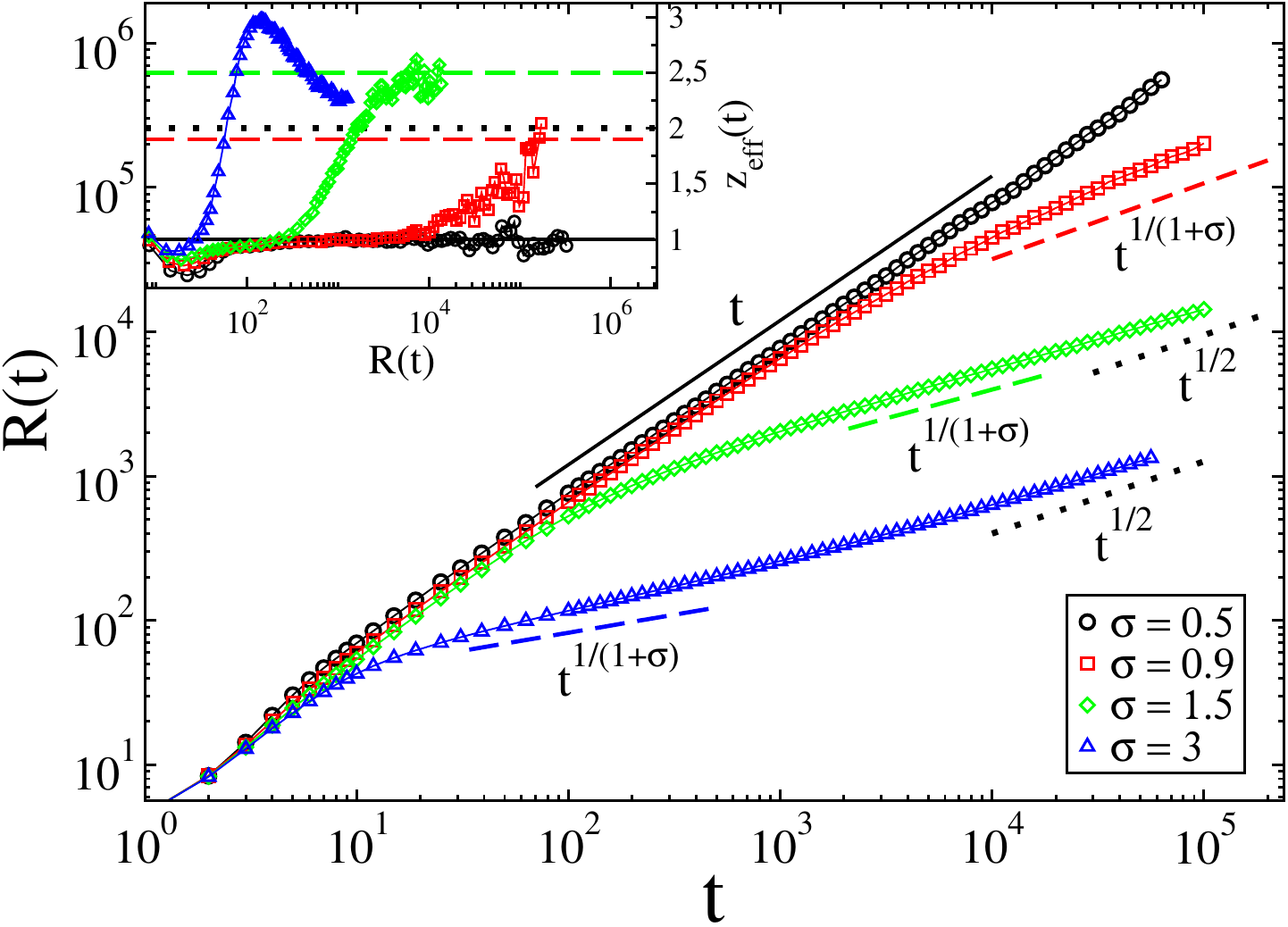}
	\caption{{\textbf{1D LRIM:} Log-log plot of $R(t)$ vs. $t$ for the $1D$ pure LRIM ($\Delta = 0$) quenched to $T = 10^{-3}$ (in units of $J_0$) with different $\sigma$-values. The inset shows the effective exponent $z_{\rm eff}$ vs. $R(t)$ for the data in the main figure. The solid, dashed and dotted  lines (also shown in the inset) denote the predicted  algebraic growth laws $R(t)\sim t^{1/\overline z}$}. Specifically, the solid black line refers to the pre-asymptotic ballistic law with $\overline z=1$; the dashed lines denote growth with $\overline z=\sigma+1$; and the dotted black line shows the diffusive behavior with $\overline z=2$, which is the asymptotic growth-law for $\sigma >1$.}
	\label{fig2}
\end{figure}

In Fig.~\ref{fig2}, we show the above growth laws for quench temperature $T = 10^{-3}$ and different $\sigma$-values. Notice that the length of pre-asymptotic regimes diminishes as $T$ or $\sigma$ increases, which is clearly observed in the figure. When $\sigma = 0.5$, a clean ballistic regime is obtained over multiple decades of time. As $\sigma$ increases, the pre-asymptotic ballistic growth becomes confined to shorter timescales (eventually negligible for large $\sigma$), and the subsequent growth regimes appear. For $\sigma = 3$, we observe an initial pre-asymptotic regime characterized by an effective dynamical exponent $z_{\mathrm{eff}} \simeq 1+\sigma$, followed at longer times by the expected diffusive behavior. This crossover is clearly reflected in the non-monotonic behavior of $z_{\mathrm{eff}}$ (inset), which gradually converges to the asymptotic value $z = 2$. As $\sigma$ approaches $1$, the pre-asymptotic regime becomes longer, as illustrated by the case $\sigma = 1.5$, where $z_{\mathrm{eff}} \simeq 1 + \sigma$ persists over a broad temporal range. We emphasize that capturing all predicted regimes in a single simulation at fixed parameters $(T, \sigma)$ remains numerically challenging, as it requires impractically large system sizes~\cite{corberi2019one}.

\begin{figure}[t]
	\centering
	\includegraphics[width=0.7\textwidth]{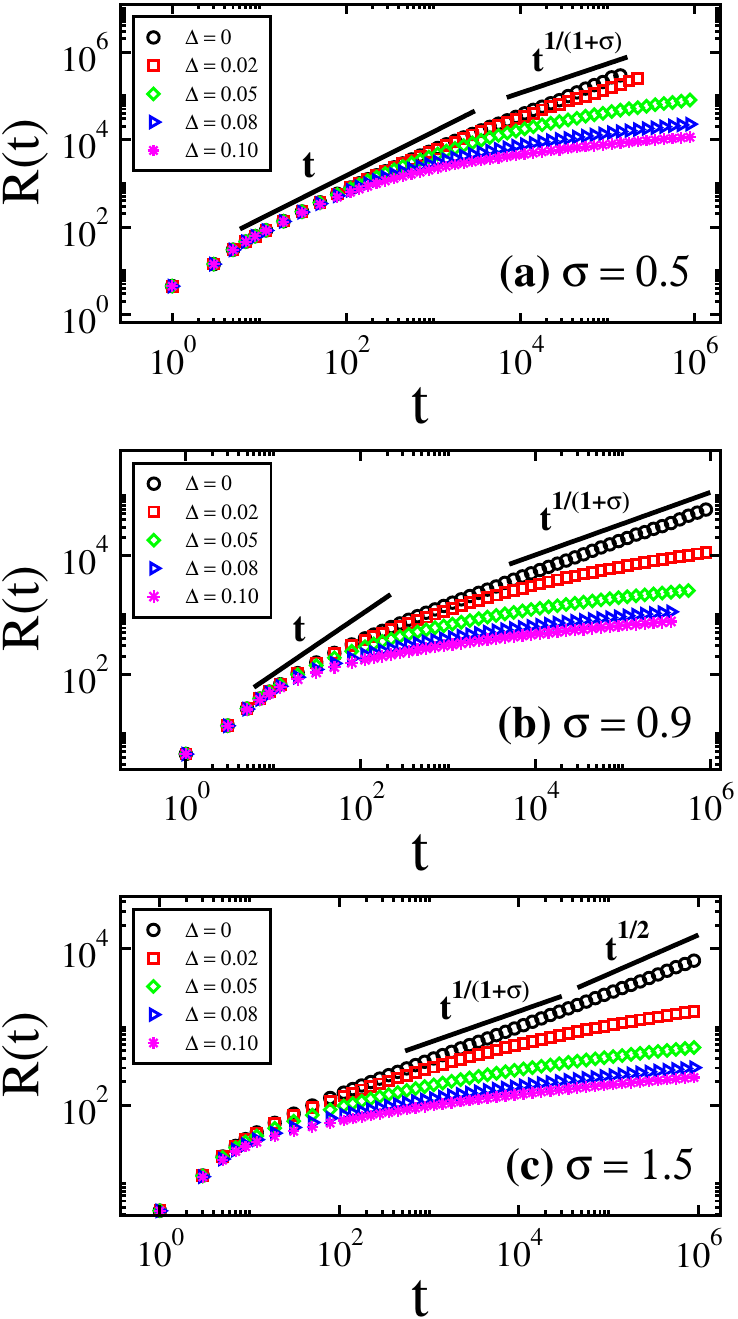}
	\caption{\textbf{1D RFLRIM:} Plot of $R(t)$ vs. $t$ for a quench to $T = 0.1$. The different panels correspond to $\sigma =$ (a) 0.5, (b) 0.9, (c) 1.5. The data sets in each panel represent different disorder values $\Delta$ (see the key). The solid lines in different frames denote the expected domain growth laws in the pure case.}
	\label{fig3}
\end{figure}

\begin{figure}[t]
	\centering
	\includegraphics[width=0.7\textwidth]{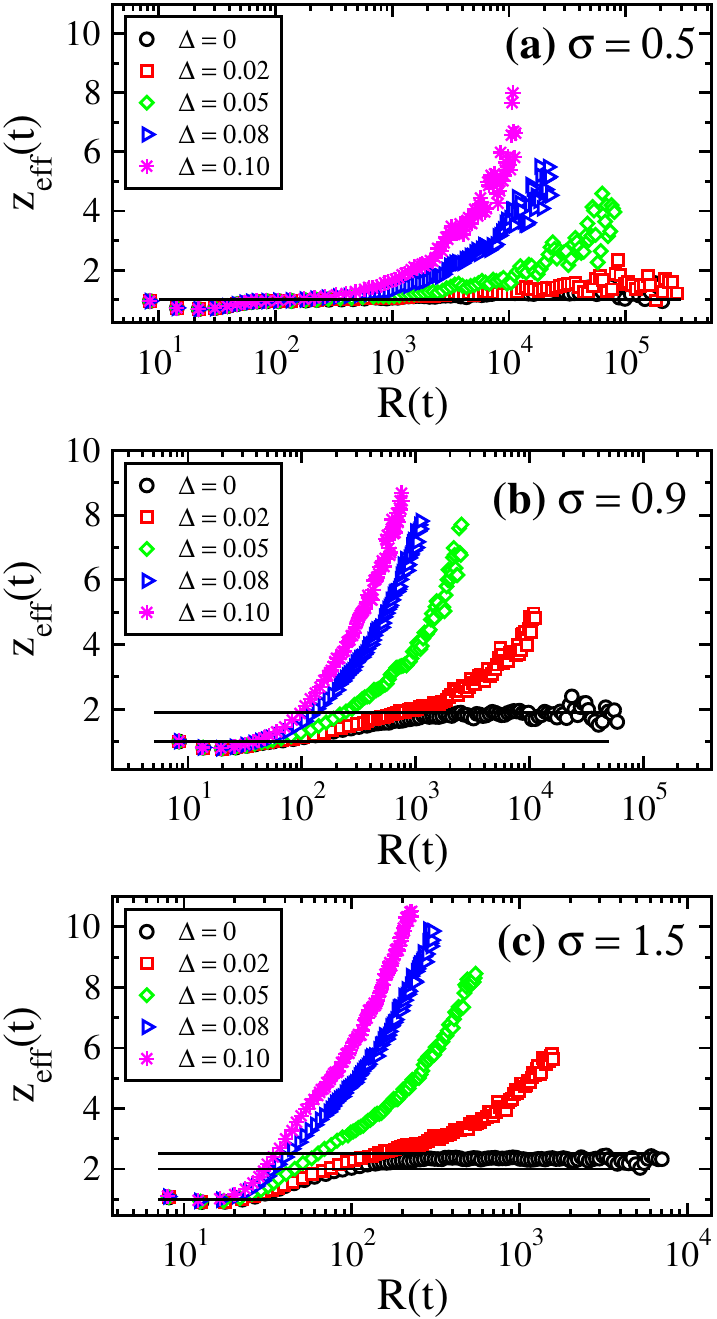}
	\caption{\textbf{1D RFLRIM:} Plot of $z_{\rm eff}(t)$ vs. $R(t)$ for the data in Fig.~\ref{fig3}. The solid lines denote the expected pre-asymptotic and asymptotic behaviors in a pure system.}
	\label{fig4}
\end{figure}

We now examine domain growth laws in the 1D RFLRIM. In Fig.~\ref{fig3}, we plot $R(t)$ vs. $t$ for different $\Delta$ and various $\sigma$-values: (a) $\sigma = 0.5$, (b) $\sigma = 0.9$, and (c) $\sigma = 1.5$. The quench temperature is set to $T = 0.1$ for all panels. The corresponding effective dynamical exponents $z_{\rm eff}(t)$ are shown in Fig.~\ref{fig4}. We choose disorder strengths $\Delta \leq T$, which is motivated by competing energy scales in disordered systems: Quenched disorder introduces energy barriers $E_{\rm B} \sim \Delta [R(t)]^{\psi}$~\cite{PhysRevB.37.9481} that are balanced by thermal energy $E_{\rm T} \sim k_{\rm B}T$, where the ratio $\Delta/T$ governs the $R$-dependence of these barriers and drives the activated growth dynamics.

We begin from Fig.~\ref{fig3}(b) for $\sigma = 0.9$. When $\Delta = 0$, the expected ballistic growth is observed on the early time scales, which then crosses over to the prediction of BR in \eqref{BR}. This is more apparent in the corresponding plot for $z_{\rm eff}$ in Fig.~\ref{fig4}(b). When $\Delta$ becomes nonzero, the $R(t)$ data initially coincide with that of $\Delta = 0$. However, at later times, a crossover to much slower growth is observed. This is also indicated by an upward trend in $z_{\rm eff}$ after an initial flat regime. For larger $\Delta$, the system enters the slow-growth regime earlier. Notice that this slow growth is a consequence of quenched disorder. Although the upward monotonic growth in quantity $z_{\rm eff}$ suggests logarithmic growth \cite{corberi2011growth,corberi2012crossover}, a more quantitative analysis is needed for definitive conclusions (see below). Similar behavior is observed for other $\sigma$-values (Figs.~\ref{fig3}-\ref{fig4}), although the specific pre-asymptotic regimes naturally depend on $\sigma$.

To quantitatively characterize crossover dynamics, we propose a working hypothesis in which growth transitions from an initial power law $R(t) \sim t^{1/\bar{z}}$ to an asymptotic logarithmic form $R(t) \sim (\ln(t/\tau))^{\alpha}$. Here, the growth exponent $\alpha$ depends on $\sigma$, while the crossover time $\tau$ depends on both disorder strength $\Delta$ and interaction range $\sigma$ -- contrasting with the NN case where $\alpha=2$ and $\tau$ depend solely on $\Delta$.

Guided by this hypothesis, we employ the scaling analysis of Lippiello et al.~\cite{corberi2011growth,corberi2012crossover,PhysRevE.104.044123} to test our data (Figs.~\ref{fig3}--\ref{fig4}). The method involves plotting the scaled effective exponent $z_{\rm eff} - \bar{z}$ against the rescaled domain size $R(t)/\ell(\Delta)$, where $\bar{z}$ is the pre-asymptotic power-law exponent, and $\ell(\Delta) \simeq \tau^{\bar{z}}$ represents the $\Delta$-dependent crossover length. The latter is obtained by enabling a data collapse for different data sets belonging to different $\Delta$ in such a plot.

\begin{figure}[t]
	\centering
	\includegraphics[width=0.9\textwidth]{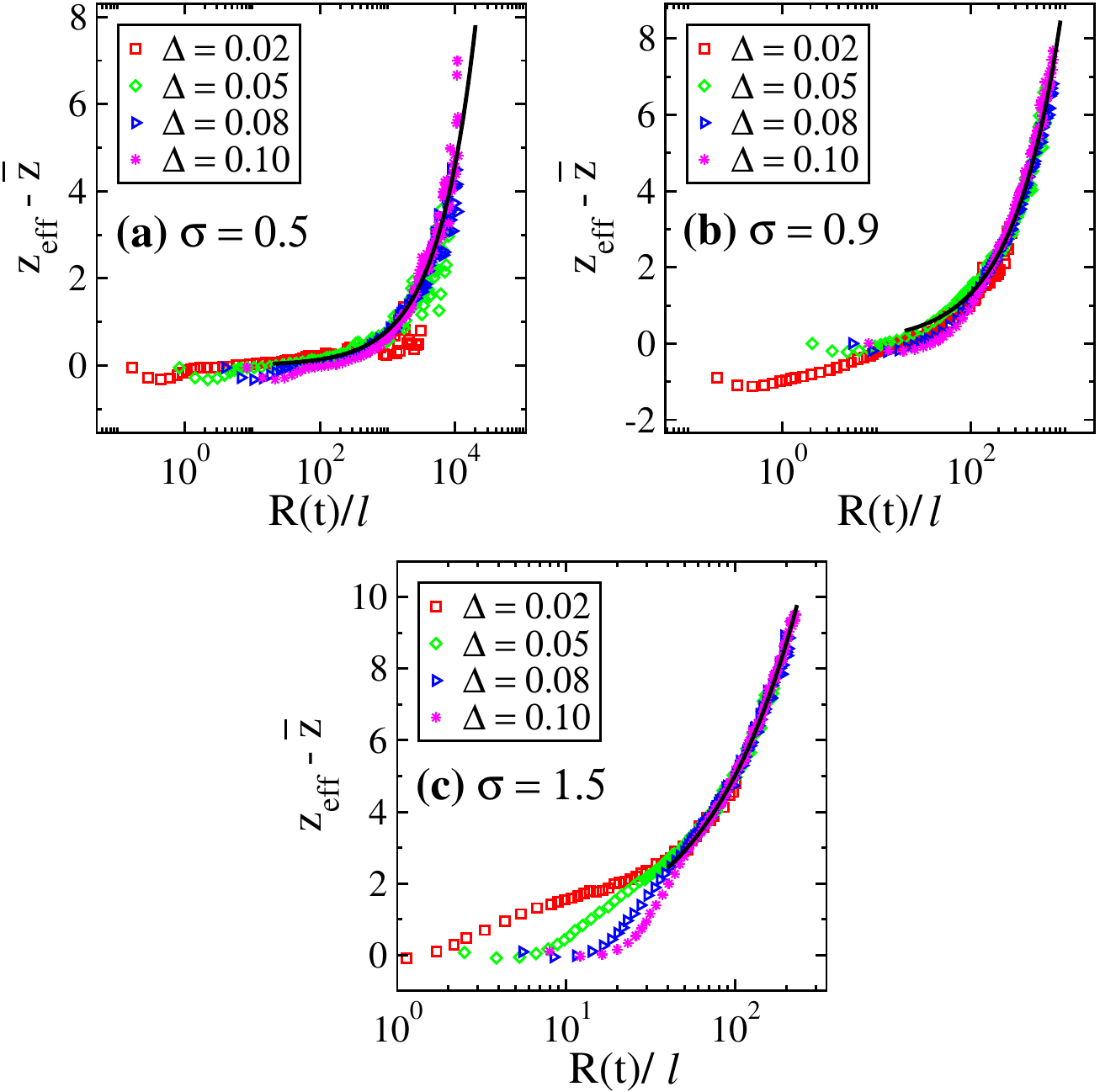}
	\caption{\textbf{1D RFLRIM:} Scaling plot of $z_{\rm eff}(t) - \bar{z}$ vs. $R(t)/\ell$ for the data in Fig.~\ref{fig4}. The solid line in different panels is the best power-law fit: $z_{\rm eff} - \bar{z} \propto (R/\ell)^{\psi}$.}
	\label{fig5}
\end{figure}

In Fig.~\ref{fig5}, the above method is exploited for the data sets in Fig.~\ref{fig4}. A nice data collapse is obtained for each value of $\sigma$, see panels (a)--(c). A power-law fit (solid line in Fig.~\ref{fig5}(a)--(c)) of the form
\be
z_{\rm eff} - \bar{z} = a(R/\ell)^{\psi (\sigma)}
\ee
to the scaled data in various panels indicates a logarithmic growth $R(t) \sim [\ln(t/\tau)]^{\alpha(\sigma)}$, where exponent $\alpha(\sigma) = 1/\psi(\sigma)$. Notice that $\psi(\sigma)$ is the barrier exponent, which characterizes the scaling of disorder-induced energy barriers as $E_{\rm B} \sim \Delta [R(t)]^{\psi}$, and thus controls the slow, activated nature of coarsening dynamics. The measured values of $\alpha(\sigma)$ are: $1.31(5)$ for $\sigma = 0.5$, $1.17(1)$ for $\sigma = 0.9$, and $1.26(1)$ for $\sigma = 1.5$. Clearly, $\alpha(\sigma)$ is less than the NN limit $\alpha(\infty) = 2$ -- a surprising result that reveals a systematic reduction in the growth efficiency in long-range disordered systems. We refer the reader to Ref.~\cite{PhysRevE.108.044131} for further details.

\begin{figure}[t]
	\centering
	\includegraphics[width=0.9\textwidth]{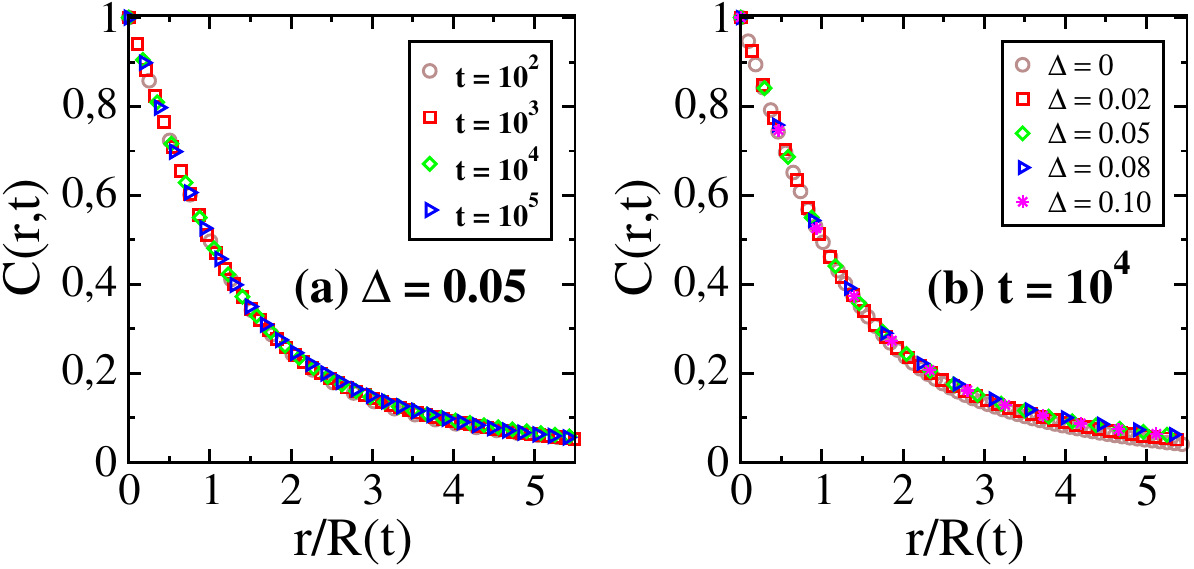}
	\caption{\textbf{1D RFLRIM:} Scaling plots of the spatial correlation function, $C(r,t)$ vs. $r/R(t)$, for a quench to $T = 0.1$ of a system with $\sigma = 0.9$. The panel (a) shows data sets at different times $t$ for a fixed disorder amplitude $\Delta = 0.05$, and (b) shows data sets at fixed $t = 10^4$ with various $\Delta$.}
	\label{fig6}
\end{figure}

For completeness, we also present the results for the dynamical scaling of the spatial correlation function $C(r,t)$; see Eqs.~\eqref{corr}-\eqref{cf_scale}. In Fig.~\ref{fig6}(a), we plot $C(r,t)$ vs. $r/R(t)$ of a system with $\sigma = 0.9$ and $\Delta = 0.05$ for various timescales in terms of MCS. Different data sets collapse onto a single scaling function, confirming the dynamical scaling signature of $C(r,t)$. In Fig.~\ref{fig6}(b), we plot $C(r,t)$ vs. $r/R(t)$ for different $\Delta$-values, at a fixed $t=10^4$ MCS. This demonstrates the super-universal (SU) nature of the scaling function for $\sigma = 0.9$, that is, the scaling function $f(x)$ (in Eq.~\eqref{cf_scale}) is independent of the disorder strength. We remark that the SU nature of the scaling function is also obeyed for other values of $\sigma > 0$, see Ref.~\cite{PhysRevE.108.044131}.

\section{Domain Growth in Two Dimensions}
\label{s5}

In this section, we present our numerical results for the $2D$ RFLRIM. The growth law $R(t)$ is calculated for a system of size $2048^2$ by averaging over 25 independent simulation runs, with each run consisting of different random initial conditions and disorder realization.

\begin{figure}[t]
	\centering
	\includegraphics[width=0.9\textwidth]{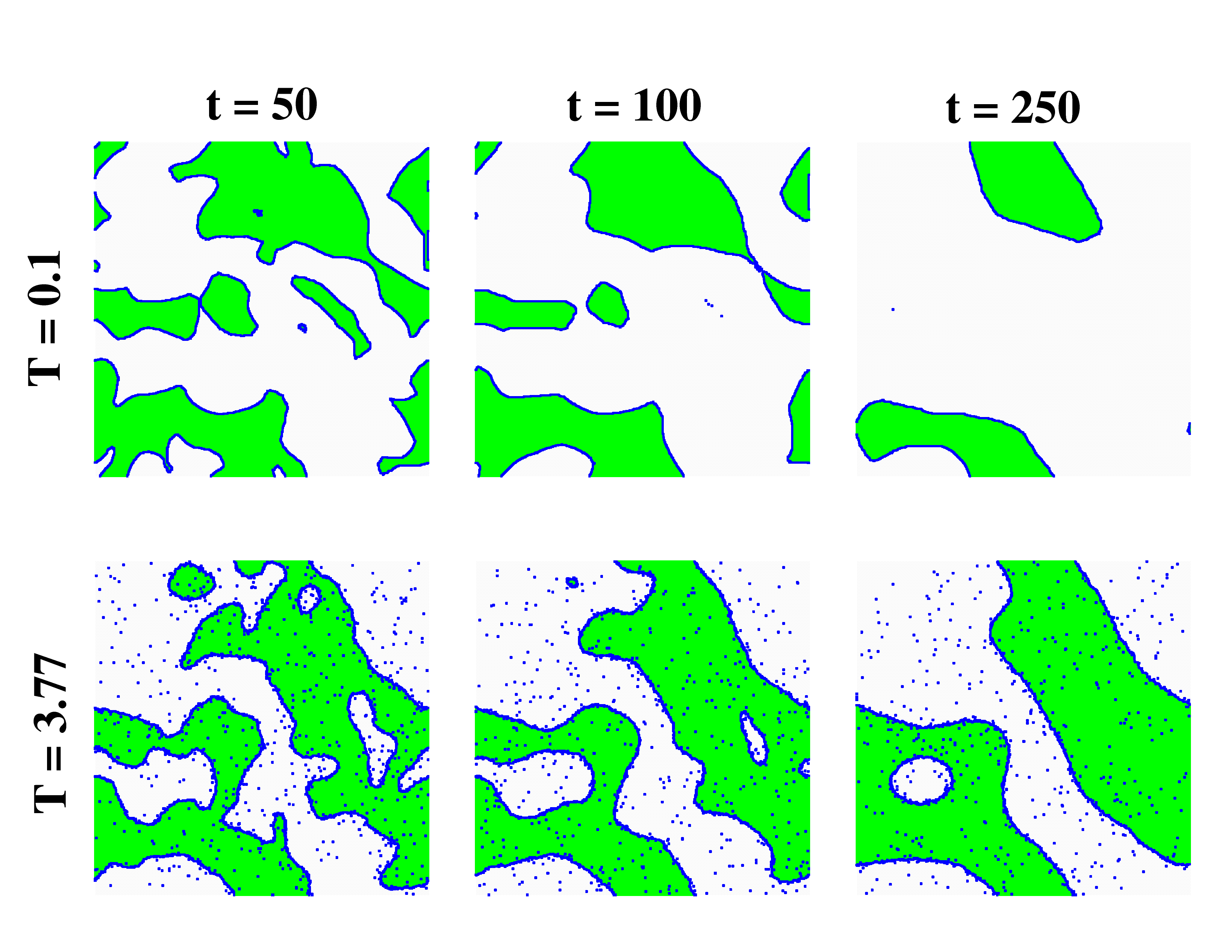}
	\caption{\textbf{2D RFLRIM:} Evolution snapshots of domain interfaces in a spin configuration for fixed disorder value $\Delta = 0.1$ at different times $t$ (see the key) after an instantaneous quench to $T = 0.1$ (upper row) and $T = 3.77$ (lower row) of a system with long-range parameter $\sigma = 0.6$. The linear size of system is fixed to $L = 512$. While interfaces are marked in blue (dark), the up spins are shown in green (light).}
	\label{fig7}
\end{figure}

The domain growth mechanics in $2D$ RFLRIM exhibits greater complexity due to the increased dimensionality. To illustrate this, Fig.~\ref{fig7} presents evolution snapshots of spin configurations for fixed $\Delta=0.1$ and $\sigma=0.6$, comparing two temperature quenches: $T=0.1$ (upper row) and $T=3.77$ (lower row). Notice that $T_c \simeq 12.55$ is the critical temperature of the pure system with $\Delta=0$ \cite{PhysRevE.95.012143}. Let us first focus on the upper row, where clean domains enriched with aligned spins (up or down) are formed at different time-steps. Moreover, as time progresses, the small domains disappear leaving behind only larger ones. The domain interfaces cost energy, and during dynamics, the system minimizes net free energy by annealing its interfaces. The picture is similar for a quench to a higher temperature (lower row). The larger domains are formed at the cost of smaller ones. However, the presence of \textit{increased} thermal agitation randomizes the motion of the interface while allowing metastable sub-domains to persist within larger ones.

\begin{figure}[t]
	\centering
	\includegraphics[width=0.9\textwidth]{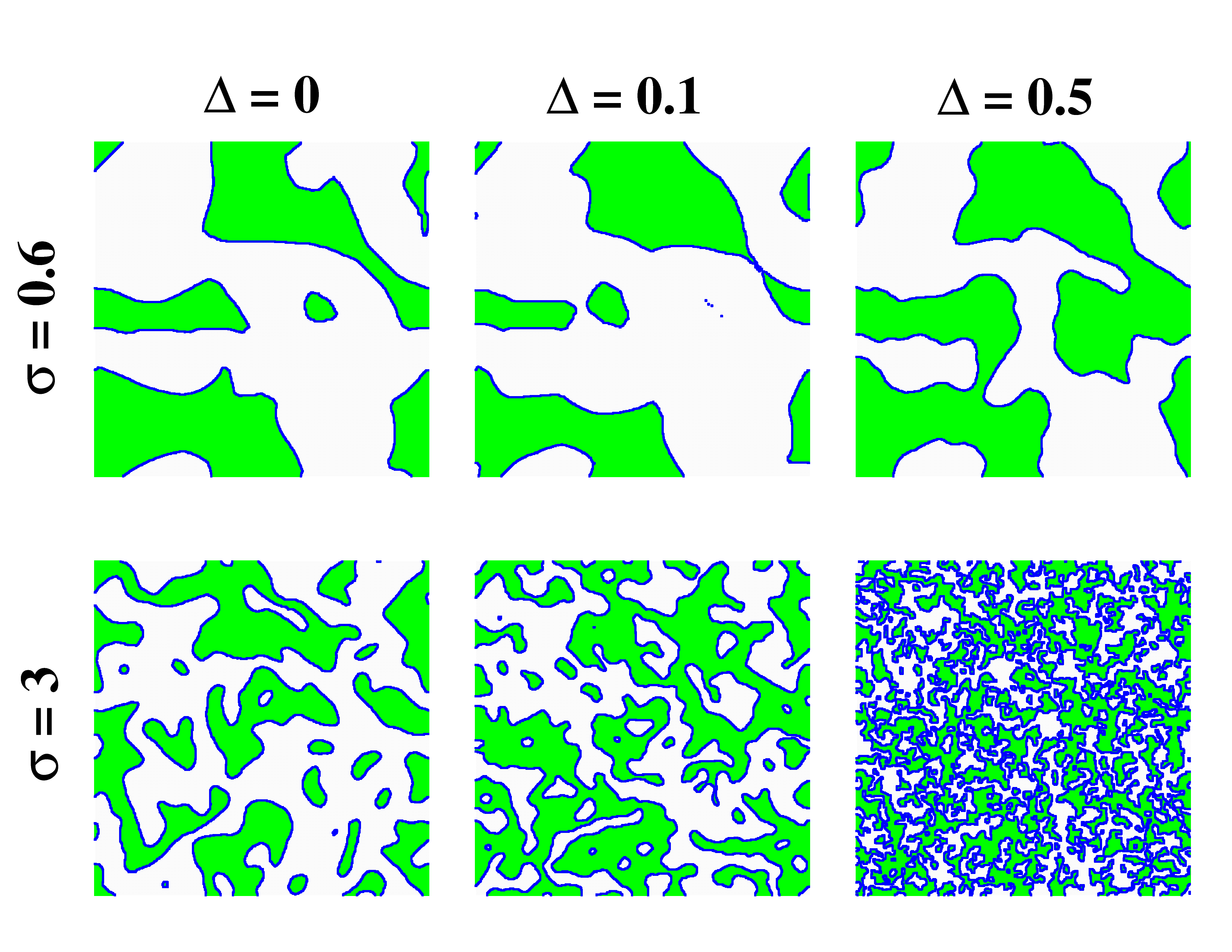}
	\caption{\textbf{2D RFLRIM:} Snapshots of domain interfaces in spin configurations for various disorder amplitudes $\Delta$ (see the key) at fixed time $t = 250$ after an instantaneous quench to $T = 0.1$ of a system with long-range parameter $\sigma = 0.6$ (upper row) and $\sigma = 3$ (lower row). The linear size of the system is fixed to $L = 512$.}
	\label{fig8}
\end{figure}

In Fig.~\ref{fig8}, we show the spin configurations at a fixed time $t = 250$ but different disorder values $\Delta = 0,0.1,0.5$, for a system with $\sigma = 0.6$ (upper row) and $\sigma = 3$ (lower row). The on-site quenched random field pins the domain interfaces in the local energy barriers (see Fig.~\ref{fig1}). Therefore, a slower growth of domains is expected with disorder, as seen in Fig.~\ref{fig8} for $\sigma = 3$. The density of the interfaces increases with the amplitude of the disorder, which indicates that growth is slower at large $\Delta$. However, for $\sigma = 0.6$, the growth appears to be less affected by the presence of disorder. This is because the strong LR interaction facilitates escape from disorder-induced barriers. Direct comparison between panels reveals that enhanced long-range coupling ($\sigma=0.6$ vs. $\sigma=3$) consistently accelerates domain growth regardless of $\Delta$.

To understand the interplay of disorder and long-range interaction quantitatively, we now turn to the domain growth law. For clarity and continuity, we begin by discussing the behavior in the pure system. As in the one-dimensional case, the BR predictions~\eqref{BR} also hold in two dimensions. In addition to these, a distinct power law (with $z = 4/3$) is observed for a zero-temperature quench~\cite{PhysRevE.103.012108,PhysRevE.103.052122,PhysRevE.105.034131}. The latter is universal with respect to $\sigma$ and is a consequence of the increased metastable features in $D=2$~\cite{PhysRevE.105.034131}. Obviously, when quenching to small but non-zero $T$, a crossover from this universal law to the BR law is observed~\cite{PhysRevE.103.012108}.

\begin{figure}[t]
	\centering
	\includegraphics[width=0.9\textwidth]{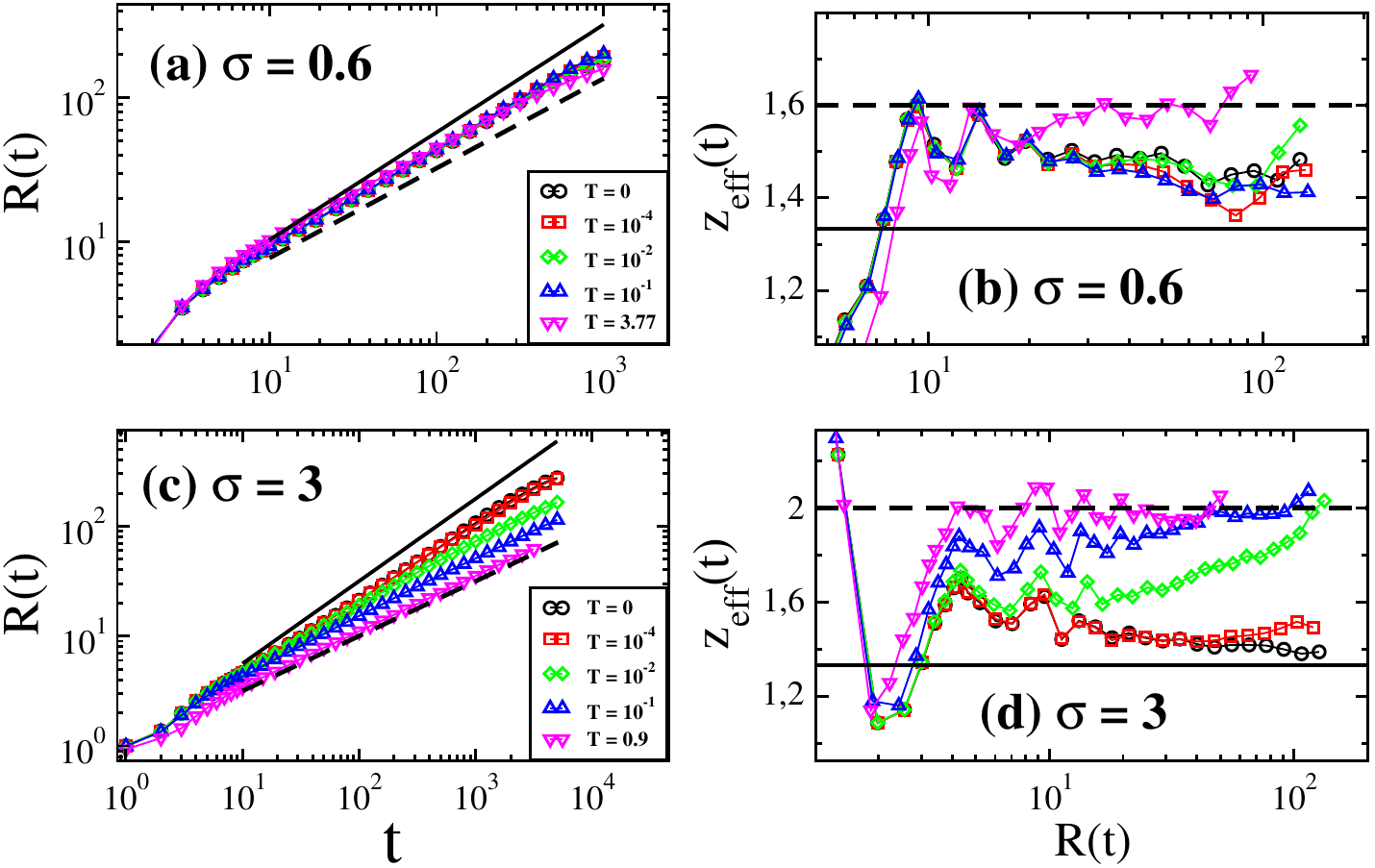}
	\caption{\textbf{2D LRIM:} Plot of $R(t)$ vs. $t$ on a log-log scale for a pure long-range system with (a) $\sigma = 0.6$ and (c) $\sigma = 3$ quenched to different temperatures (see the key). The values of $T_c$ are \cite{PhysRevE.95.012143}: $T_c(\sigma = 0.6) \simeq 12.55$, $T_c(\sigma = 3) \simeq 3.0$. The panels (b) and (d) plot the effective exponent $z_{\rm eff}$ against $R(t)$, on a log-linear scale, for data in (a) and (c), respectively. The solid and dashed lines denote the expected pre-asymptotic ($\bar{z} = 4/3$) and asymptotic ($\bar{z} = 1.6$ for $\sigma = 0.6$ and $\bar{z} = 2$ for $\sigma = 3$) behavior, respectively.}
	\label{fig9}
\end{figure}

In Fig.~\ref{fig9}, we compare the above predictions for $\sigma = 0.6, 3$ and different $T$. First, let us focus on the case with $\sigma =3$ in (c)-(d). For a quench to a relatively higher temperature ($T = 0.3T_c$), the dynamical exponent $z_{\rm eff}$ asymptotically reaches the BR exponent $z = 2$, while for a quench to zero $T$ it approaches $z = 4/3$. When the system is quenched to a low but non-zero $T$, a transitional behavior in $R(t)$ is obtained. A similar pattern is observed for $\sigma = 0.6$.

\begin{figure}[t]
	\centering
	\includegraphics[width=0.9\textwidth]{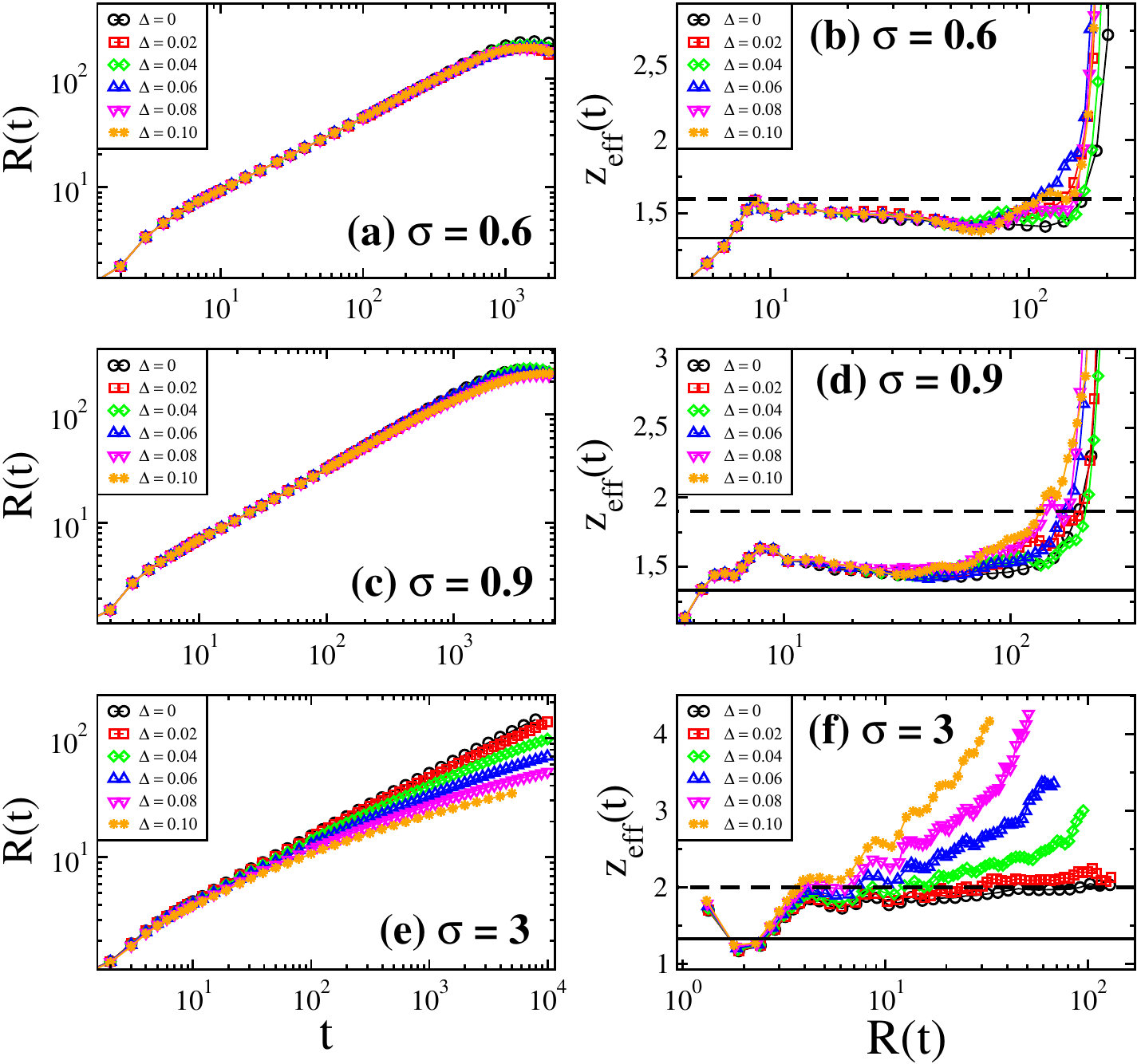}
	\caption{\textbf{2D RFLRIM:} Plot of $R(t)$ vs. $t$ on a log-log scale for system with (a) $\sigma = 0.6$, (c) $\sigma = 0.9$, and (e) $\sigma = 3$, quenched to $T=0.1$. The panels (b), (d), and (f) plot the effective exponent $z_{\rm eff}$ against $R(t)$, on a log-linear scale, for data in (a), (c), and (e), respectively. Different data sets in each panel belong to different $\Delta$-values (see the key). The solid and dashed lines denote the expected pre-asymptotic ($\bar{z} = 4/3$) and asymptotic ($\bar{z} = 1+\sigma$ for $\sigma < 1$ and $\bar{z} = 2$ for $\sigma > 1$) behavior in a pure system, respectively.}
	\label{fig10}
\end{figure}

Now we discuss our results for domain growth in the disordered case. In Fig.~\ref{fig10}, we investigate the growth law for $\sigma = 0.6, 0.9, 3$. The quench temperature is fixed at $T = 0.1$, and the disorder amplitude $\Delta$ is chosen such that $\Delta \le T$ (similar to the $1D$ case Let us start from panels (a)-(b) of $\sigma = 0.6$ in Fig.~\ref{fig10}. For all $\Delta$-values (including $\Delta=0$) in the plots, $R(t)$ remains $\Delta$-independent up to $R(t) \leq 100$. In all cases, the growth law initially follows the BR regime, characterized by $z_{\text{eff}} \simeq 1 + \sigma$, and gradually approaches the asymptotic value $z = 4/3$. The subsequent divergence of $z_{\rm eff}$ for $R(t)>100$ probably reflects finite-size effects rather than true dynamical freezing, since even the pure system ($\Delta=0$) shows unimpeded growth up to $R(t)\leq100$ at the current system size ($L=2048$). Similar behavior occurs for $\sigma=0.9$ [Figs.~\ref{fig10}(c)-(d)], where $\Delta$-independent growth persists until $R(t)\approx100$ before $z_{\rm eff}$ diverges. These observations demonstrate that long-range interactions with small $\sigma$ dominate over disorder-induced pinning at intermediate scales ($R(t)\leq100$).

In panels (e)-(f) of Fig.~\ref{fig10}, the growth law for $\sigma = 3$ is analyzed, which reveals a quite distinct picture. When $\Delta = 0$, $R(t)$ asymptotically agrees with the BR prediction $R(t) \sim t^{1/\bar{z}}$, with $\bar{z} = 2$. As $\Delta$ begins to increase, the growth of $R(t)$ is no longer a power law and shows considerable differences from the pure case; see the deviation of $R(t)$ from a straight line on the log-log plot in panel (e) and the smooth upward increase in $z_{\rm eff}$ with $R(t)$. For large values of $\Delta$, the growth of $R(t)$ is slower. This transition to the slow regime mirrors the behavior observed in NN ($\sigma \rightarrow \infty$) system~\cite{corberi2012crossover}. Recall that at large $\Delta$, $R(t) < 100$ in Figs.~\ref{fig10}(e)-(f).

\begin{figure}[t]
	\centering
	\includegraphics[width=0.9\textwidth]{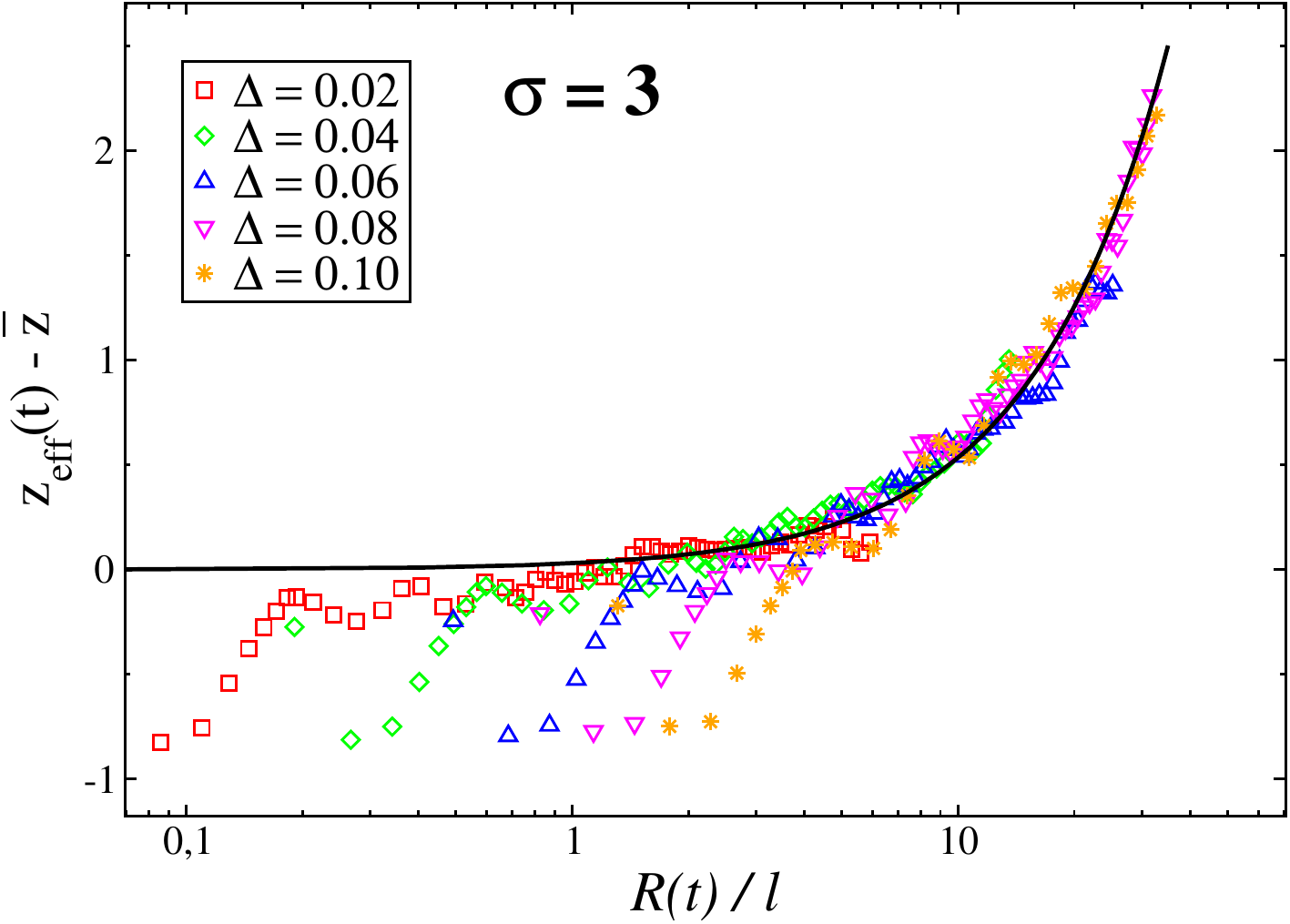}
	\caption{\textbf{2D RFLRIM:} Scaling plot of $z_{\rm eff}(t) - \bar{z}$ vs. $R(t)/\ell$ for the data in Fig.~\ref{fig10}(f) of $\sigma = 3$. The solid line is the best power-law fit: $z_{\rm eff} - \bar{z} \propto (R/\ell)^{\psi}$, with $\psi \simeq 1.23$.}
	\label{fig11}
\end{figure}

To verify whether the above could be a signature of expected logarithmic growth, we again use the method of Lippiello et al. \cite{corberi2011growth,corberi2012crossover,PhysRevE.104.044123} as described in Sec.~\ref{s4}. For this purpose, in Fig.~\ref{fig11} we plot the scaled value of effective exponent $z_{\rm eff} - \bar{z}$ against $R(t)/\ell (\Delta)$, where pre-asymptotic exponent $\bar{z}$ is set to $\bar{z} = 2$, and the crossover length $\ell (\Delta)$ is chosen so as to obtain a data collapse for different data sets in the figure. For small values of $R(t)/\ell < 4$, no data collapse is observed, which was expected since this regime corresponds to the pre-asymptotic growth of $R(t)$ (see Fig.~\ref{fig10}(f)). For large $R(t)/\ell$, a clean scaling collapse is obtained for different data sets in the figure. The power-law fit of the form $z_{\rm eff} - \bar{z} = a(R/\ell)^{\psi}$ to the collapse gives an exponent $\psi \simeq 1.23$. For comparison, the corresponding value in the NN case is $\psi \simeq 1.5$, indicating only a modest difference. This confirms logarithmic growth for $\sigma=3$, demonstrating that when long-range interactions weaken sufficiently ($\sigma > 1$), the system crosses over to disorder-dominated activated dynamics.

\begin{figure}[t]
	\centering
	\includegraphics[width=0.9\textwidth]{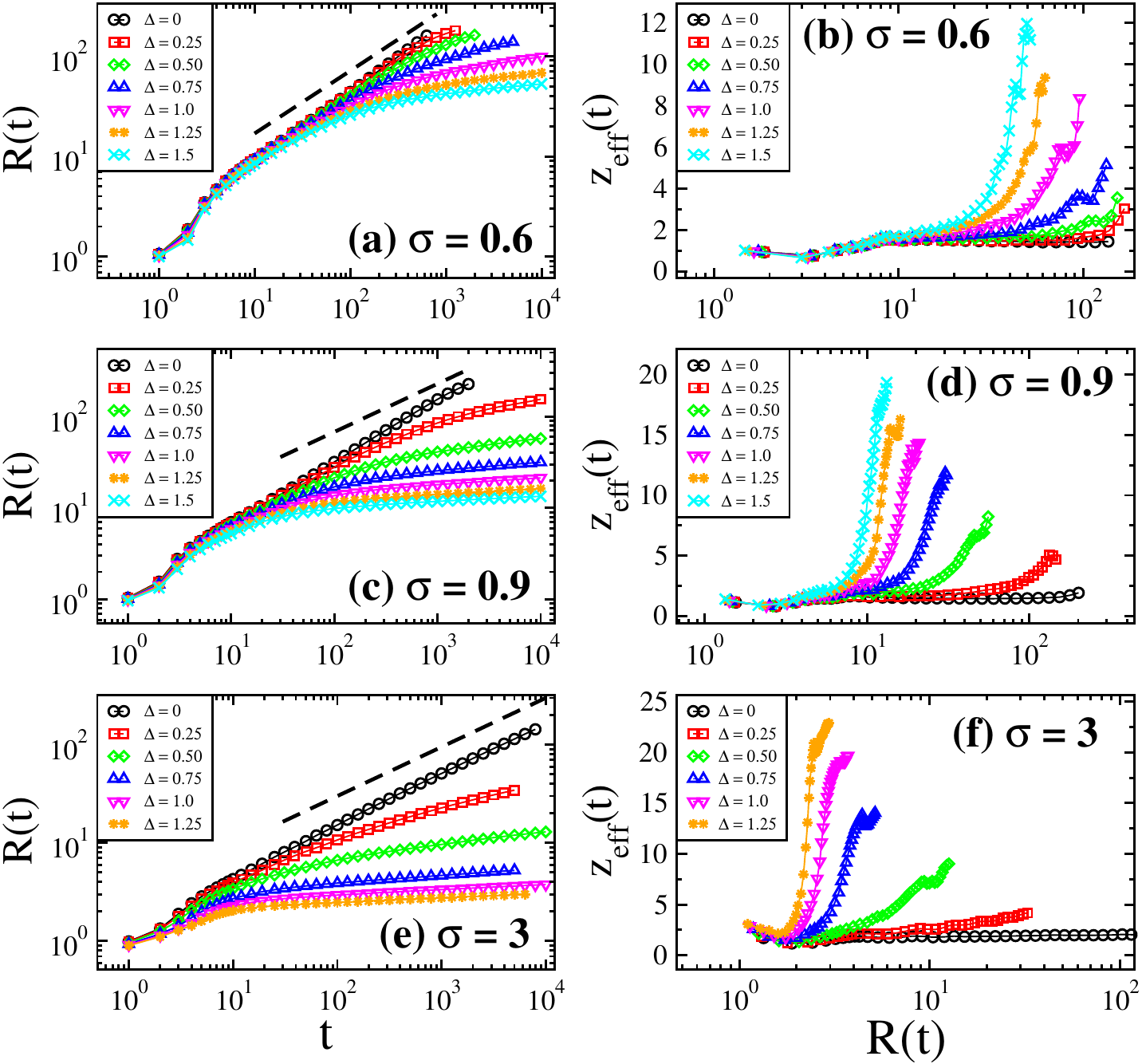}
	\caption{\textbf{2D RFLRIM:} Plots similar to Fig.~\ref{fig10} but for large values of disorder amplitude $\Delta > T$ and $\Delta = 0$. The dashed lines denote the expected asymptotic behavior ($R(t) \sim t^{1/{\bar{z}}}$, with $\bar{z} = 1+\sigma$ for $\sigma < 1$ and $\bar{z} = 2$ for $\sigma > 1$) in a pure system, respectively.}
	\label{fig12}
\end{figure}

To examine the behavior at small values of $\sigma$ more sensitively, we further analyze domain growth in the regime of strong disorder where $\Delta > T$. Similarly to Fig.~\ref{fig10}, Fig.~\ref{fig12} presents data for $\sigma = 0.6, 0.9, 3$ at the same $T=0.1$, but with different disorder amplitudes set to $\Delta > T$. For comparison, the pure case ($\Delta=0$) is also included. We begin with panels (a)-(b) of $\sigma = 0.6$. Unlike the small $\Delta$ values in Fig.~\ref{fig10}, the growth law $R(t)$ changes considerably with $\Delta$. When $R(t)$ is small, the growth remains identical to the pure case. Beyond a crossover value of $R(t)$ which decreases substantially with $\Delta$, the growth law is completely different from a power law: an extremely slow growth regime is obtained, indicated by a pronounced plateau in $R(t)$ and a sharp increase in the dynamical exponent $z_{\rm eff}$. It is important to note that the slow growth observed here is not due to finite-size effects. Compared to the logarithmic growth regime observed in Fig.~\ref{fig10}(f), the increase of $z_{\rm eff}$ with $R$ in Fig.~\ref{fig12}(b) is much more rapid. Motivated by this, we attempted a scaling collapse following the framework used to identify logarithmic growth, but no collapse could be obtained -- suggesting instead a dynamical freezing. Similar trends are observed for $\sigma=0.9, 3$ in Figs.~\ref{fig12}(c)-(f).

In conclusion, when the disorder amplitude $\Delta$ is large and the quench temperature $T$ is small, the system enters an extremely slow dynamical regime. This slowdown arises because the energy barriers induced by disorder dominate over the thermal energy required for activation. Since the barrier height scales as $E_{\rm B} \sim \Delta [R(t)]^{\psi}$, large $\Delta$ amplifies these barriers as the domain size grows, significantly impeding further coarsening.

\begin{figure}[t]
	\centering
	\includegraphics[width=0.9\textwidth]{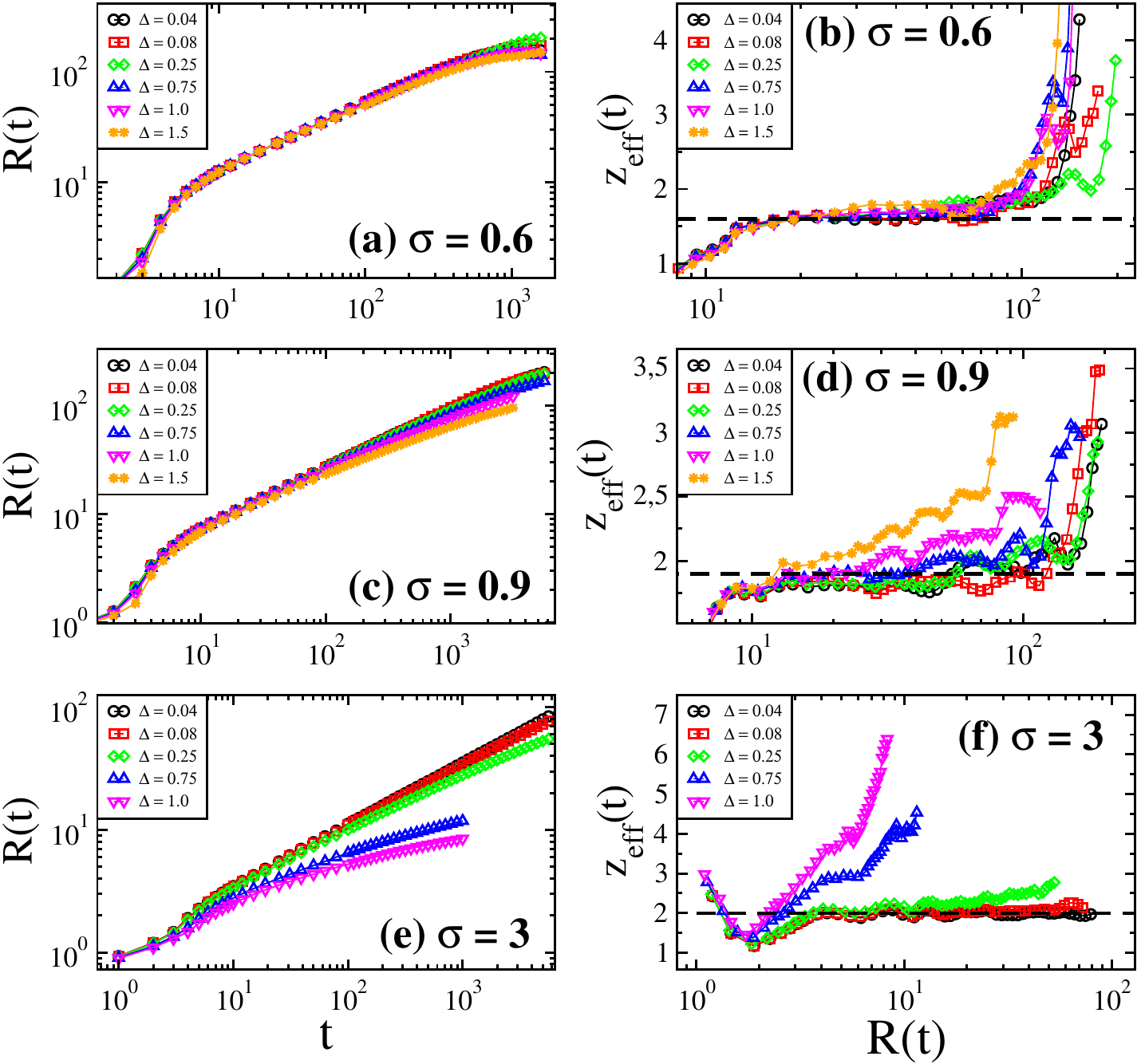}
	\caption{\textbf{2D RFLRIM:} Plot of $R(t)$ vs. $t$ on a log-log scale for system with (a) $\sigma = 0.6$, (c) $\sigma = 0.9$, and (e) $\sigma = 3$, quenched to relatively high temperatures $T = 0.3 T_c(\sigma)$. The panels (b), (d), and (f) plot the effective exponent $z_{\rm eff}$ against $R(t)$, on a log-linear scale, for data in (a), (c), and (e), respectively. Different data sets in each panel belong to different $\Delta$-values (see the key).}
	\label{fig13}
\end{figure}

To further deepen our understanding of the dynamics, we now present some results for a higher quench temperature, approximately $T = 0.3 T_c$, as shown in Fig.~\ref{fig13}. Here, $T_c$ denotes the critical temperature of the pure case and is extracted from Ref.~\cite{PhysRevE.95.012143}: $T_c \simeq 12.55$ for $\sigma = 0.6$, $T_c \simeq 8.8$ for $\sigma = 0.9$, and $T_c \simeq 3.0$ for $\sigma = 3$. The $\Delta$-values are chosen so that $\Delta \leq T$. Starting with panels (a)-(b) of $\sigma = 0.6$, again there is no noticeable dependence on $\Delta$ up to $R(t) \leq 100$. Beyond $R(t) > 100$, the exponent $z_{\rm eff}$ increases sharply, but this regime cannot be excluded from finite size effects. Let us now move on to other $\sigma$-values. For $\sigma = 0.9$, the growth law changes gradually as $\Delta$ increases from small to large values; see Figs.~\ref{fig13}(c)-(d). For small $\Delta$, the growth has power law behavior (plateau in $z_{\rm eff}$), but for large $\Delta$ the $z_{\rm eff}$ changes monotonically, apart from a sudden jump at large $R$. The data for $\sigma = 3$ cleanly demonstrate a power law for small $\Delta$. As $\Delta$ increases, the growth starts to slow down, see Figs.~\ref{fig13}(e)-(f). Notice that we did not see any abrupt changes in $z_{\rm eff}$ for $\sigma = 3$.

\begin{figure}[t]
	\centering
	\includegraphics[width=0.9\textwidth]{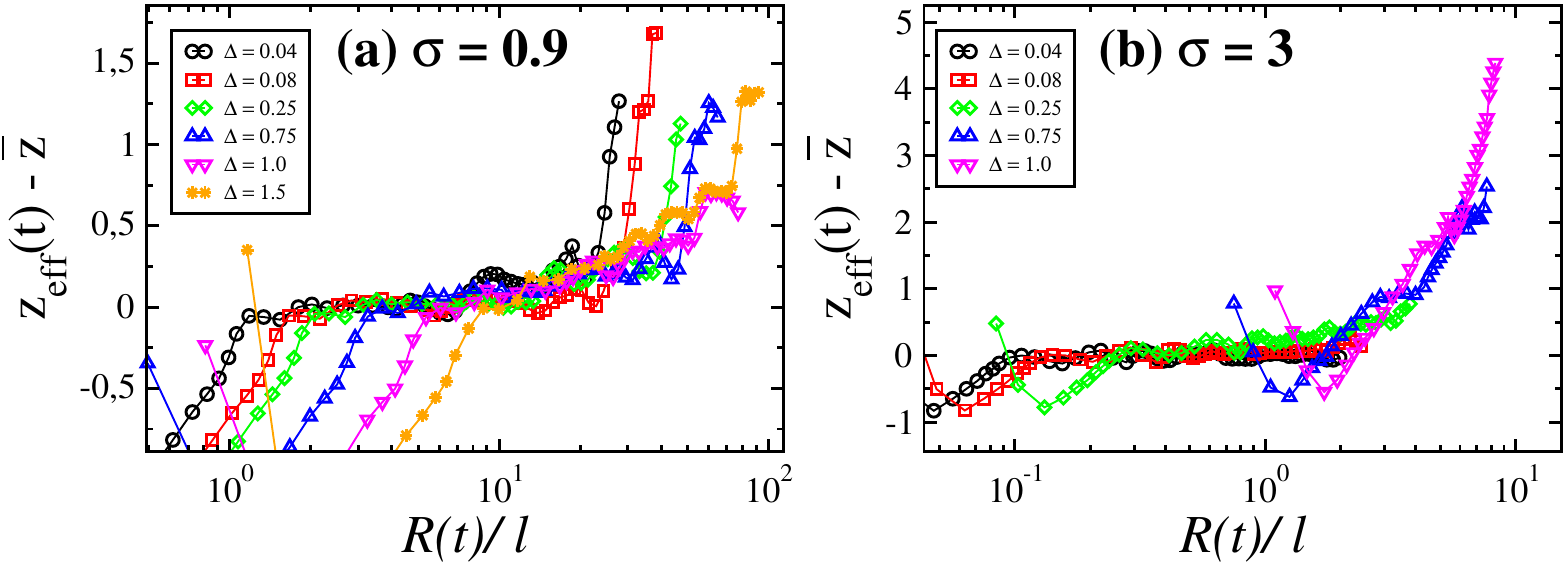}
	\caption{\textbf{2D RFLRIM:} Scaling plots of $z_{\rm eff}(t) - \bar{z}$ vs. $R(t)/\ell$ for the data in (a) Fig.~\ref{fig13}(d) of $\sigma = 0.9$ and (b) Fig.~\ref{fig13}(f) of $\sigma = 3$. The values of $\bar{z}$ for $\sigma = 0.9, 3$ are $\bar{z} \simeq 1.8, 2$, respectively.}
	\label{fig14}
\end{figure}

To further quantify these behaviors, we analyze the scaled exponent $z_{\rm eff}(t) - \bar{z}$ versus the scaled length scale $R(t)/\ell$ for the cases $\sigma = 0.9$ and $\sigma = 3$, shown in Figs.~\ref{fig14}(a) and (b). For $\sigma = 3$, the data sets exhibit good collapse, although the region where a power law fit could hold is limited to a narrow range beyond $R(t)/\ell \simeq 2$. For $\sigma = 0.9$, a partial collapse is observed only on intermediate scales, with deviations at both smaller and larger values of $R(t)$. Finally, no conclusive evidence of asymptotic logarithmic growth can be established for $\sigma < 1$ within the time and system sizes accessible in this study.

\begin{figure}[t]
	\centering
	\includegraphics[width=0.99\textwidth]{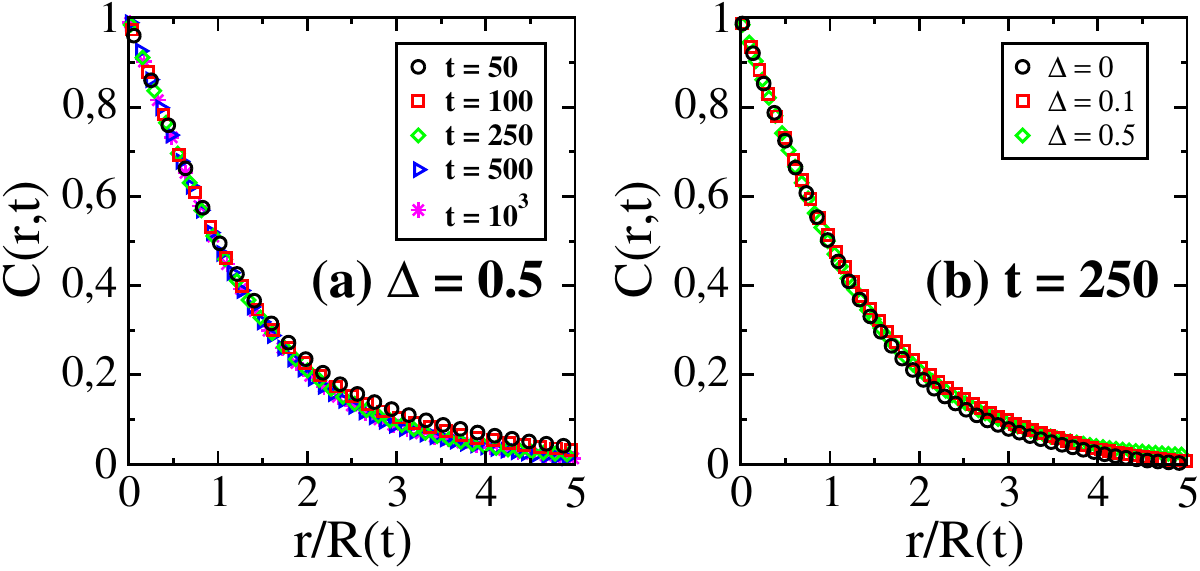}
	\caption{\textbf{2D RFLRIM:} Scaling plots of the spatial correlation function, $C(r,t)$ vs. $r/R(t)$, for a quench to $T = 0.1$ of a system with $\sigma = 0.6$. The panel (a) shows data sets at different times $t$ for a fixed disorder amplitude $\Delta = 0.5$, and (b) shows data sets at fixed $t = 250$ with various $\Delta$.}
	\label{fig15}
\end{figure}

Let us also explore the dynamical scaling of the spatial correlation function $C(r,t)$, an important feature of the domain growth kinetics. Fig.~\ref{fig15} (a) shows an excellent collapse of $C(r,t)$ versus $r/R(t)$ for $\sigma=0.6$ and $\Delta=0.5$ at different times, confirming dynamical scaling. More importantly, Fig.~\ref{fig15}(b) reveals that the scaling function remains $\Delta$-independent at fixed $t=250$, exhibiting SU -- a property we verify holds for all $\sigma>0$ studied (data not shown for brevity). This SU behavior suggests that while the strength of the disorder $\Delta$ controls the growth law $R(t)$, it does not affect the statistical morphology of the domain structure when properly scaled by $R(t)$.

\section{Summary and Discussion}
\label{s6}

This review provides a comprehensive investigation of domain growth kinetics in the random field long-range Ising model (RFLRIM) following quenches to low temperatures in one and two spatial dimensions. The primary focus is on the interplay between quenched disorder, introduced via on-site random fields, and the long-range interactions that decay with distance as a power law $J(r) \sim r^{-(D+\sigma)}$. Our results significantly extend the understanding of nonequilibrium dynamics in systems where both disorder and extended interactions are present -- a regime previously unexplored.

In one dimension (1D), our results demonstrate that the asymptotic domain growth law is logarithmic for all values of $\sigma > 0$, consistent with the Huse-Henley scenario. The presence of long-range interactions modifies the energy landscape through drift-induced mechanisms, yet it does not alter the activated nature of the coarsening at late times. The extracted growth exponents $\alpha$ at different $\sigma$ are smaller than the short-range limit $\alpha = 2$, highlighting a more nuanced interplay between disorder and long-range drift compared to the NN model; see Ref.~\cite{PhysRevE.108.044131} for details.

In two dimensions (2D), the situation is considerably richer and more delicate due to the increased complexity of interface motion and the interplay between thermal fluctuations, quenched disorder, and long-range interactions. For small disorder amplitudes and low temperature quench, specifically when $\Delta \leq T$, we observe that the presence of long-range interactions with small values of the power-law exponent (e.g., $\sigma = 0.6$) \textit{effectively} suppresses the influence of disorder. In this regime, the system maintains a transient power-law growth for domain size $R(t)$, which remains largely insensitive to disorder strength up to length scales $R(t) \leq 100$ in a system of linear size $L=2048$. Whether a true logarithmic growth sets in at larger length scales (say, up to $R(t) \sim 500$) remains an open question, underscoring the need for further investigation into the long-time dynamics. This would require accessing large systems up to $L=10^4$, which remains inaccessible to our current computational scope.

Keeping $\Delta \leq T$, an increase in $\sigma$ allows a clear dynamical crossover to emerge within accessible length scales. The growth law changes from an initial power-law regime to significantly slower kinetics, indicating the onset of disorder-dominated activated processes. This effect is most clearly seen in the case of $\sigma = 3$, where the long-range interaction \textit{effectively} mimics short-range behavior. In this regime, our scaling analysis of the effective dynamical exponent $z_{\rm eff}(t)$ provides compelling evidence for an asymptotic logarithmic growth law, consistent with theoretical expectations.

Furthermore, when the disorder amplitude exceeds the thermal energy scale ($\Delta > T$), an \textit{extremely} slower growth regime emerges across all $\sigma > 0$, characterized by a sharp increase in the effective dynamical exponent $z_{\rm eff}(t)$ with respect to $R(t)$, indicative of dynamical freezing.

Across all studied cases, the spatial correlation function $C(r,t)$ exhibits excellent dynamical scaling, and more notably, the scaling function shows super-universal (SU) behavior, remaining invariant across disorder amplitudes. This indicates that while the quenched disorder slows the growth of the domains, it does not significantly alter the morphology of the domains when the lengths are rescaled by $R(t)$.

As a direction for future research, it would be highly valuable to develop a theoretical framework that not only elucidates the origin and $\sigma$ dependence of the barrier exponent $\psi$, but also allows for a quantitative determination of disorder-induced energy barriers in disordered LR systems. It would also be valuable to investigate two-time quantities (e.g. autocorrelation function, response function) to better highlight the aging dynamics.

\subsection*{Author Contributions}

Ramgopal Agrawal, Federico Corberi, Eugenio Lippiello and Sanjay Puri: equal contributions to conceptualization, methodology, formal analysis and investigation, resources and data curation, writing. Sanjay Puri: funding acquisition.

\subsection*{Data Availability}

Data sets for various statistical quantities are available from the authors on reasonable request.

\newpage
\bibliography{biblio}

\newpage

\backmatter

\end{document}